# Technical Development of Two-Photon Optogenetic Stimulation and Its Potential Application to Brain-Machine Interfaces

Riichiro Hira[1,2,*], Yoshikazu Isomura[1]

**Affiliates:**

1. Department of Physiology and Cell Biology, Graduate School of Medical and Dental Sciences, Tokyo Medical and Dental University, Tokyo, Japan
2. High Performance Artificial Intelligent System Research Team, Center for Computational Science, RIKEN, Saitama 103-0027, Japan

***Correspondence:**

Riichiro Hira, M.D. Ph.D. rhira.phy2@tmd.ac.jp

Department of Physiology and Cell Biology, Graduate School of Medical and Dental Sciences, Tokyo Medical and Dental University, 1-5-45 Yushima, Bunkyo-ku, Tokyo, 113-8510, Japan.

**Acknowledgements:**

We thank Dr. K. Isobe at RIKEN for discussion. This work was supported by JP22wm0525007 (RH), JP19dm0207089 (YI) from AMED, JP22H02731 (RH), JP20K22678 (RH), JP21B304 (RH), JP21H05134 (RH), JP21H05135 (RH), JP21H0524 2(YI), and JP23H02589 (YI) from MEXT/JSPS, JPMJCR1751 (YI) from JST, Nakatani Foundation (RH), Shimadzu Foundation (RH), Takeda Science Foundation (RH), The Precise Measurement Technology Promotion Foundation (RH), Tateishi Science and Technology Foundation (RH), and Research Foundation for OptoScience and Technology (RH).

**Author contributions:**

R.H. wrote the manuscript. Y.I supervised the project.

# Abstract

Over the past decade, techniques enabling bidirectional modulation of neuronal activity with single cell precision have rapidly advanced in the form of two-photon optogenetic stimulation. Unlike conventional electrophysiological approaches or one-photon optogenetics, which inevitably activate many neurons surrounding the target, two-photon optogenetics can drive hundreds of specifically targeted neurons simultaneously, with stimulation patterns that can be flexibly and rapidly reconfigured. In this review, we trace the development of two-photon optogenetic stimulation, focusing on its progression toward implementations in large field of view two-photon microscopes capable of targeted multi neuron control. We highlight three principal strategies: spiral scanning, temporal focusing, and three-dimensional computer-generated holography (3D-CGH), along with their combinations, which together provide powerful tools for causal interrogation of neural circuits and behavior. Finally, we discuss the integration of these optical technologies into brain machine interfaces (BMIs), emphasizing both their transformative potential and the technical challenges that must be addressed to realize their broader impact.

# 1. Introduction

Traditionally, methods for stimulating neurons in the living brain have relied on electrophysiological techniques. For example, extracellular electrical stimulation using electrodes is still employed to probe the function of specific brain regions, but it nonspecifically activates cells and axons located near the electrode tip (**Fig. 1A**). Whole-cell patch clamp, by forming a gigaohm seal between the glass electrode and the cell membrane, prevents current leakage and allows stimulation to be strictly confined to a single targeted neuron (**Fig. 1B**). However, performing whole-cell patch clamp in vivo is technically challenging, and it is virtually impossible to apply this method to more than a few neurons simultaneously.

Optogenetic stimulation with one-photon illumination enables finer control by restricting opsin expression to specific cell types and limiting the illumination area (**Fig. 1C**). Nonetheless, because of the nature of one-photon excitation, stimulation cannot be confined to a single targeted cell, and arbitrary combinations of cells cannot be addressed. Two-photon optogenetic stimulation overcomes this limitation. Due to the nonlinearity of two-photon absorption, it is possible to stimulate targeted ensembles of cells, and this can be achieved for arbitrary combinations (**Fig. 1D**). Such arbitrary spatiotemporal stimulation patterns cannot be realized by any method other than two-photon optogenetics, which explains why this unique technique has rapidly advanced over the past fifteen years.

If arbitrary combinations of cells can be stimulated with two-photon light, what new possibilities emerge for brain–machine interfaces (BMIs)? One can envision a system in which two-photon calcium imaging is used to monitor neural activity in real time, the recorded signals are fed into an artificial intelligence algorithm, and the AI then determines which combination of cells should be stimulated next via two-photon optogenetics (**Fig. 1E**). The resulting activity is again measured by two-photon calcium imaging, completing a closed-loop interaction. If such closed-loop control could be implemented at scale and speed, BMIs would be transformed into something entirely new—not merely reading information from the brain, but enabling the brain and AI to work as a unified system to augment brain function. In this article, we provide a detailed survey of two-photon optogenetic stimulation methods, which are expected to underpin such advanced applications.

## 1-1. Optogenetics and Two-Photon Excitation

Two-photon excitation is essentially absent in the natural environment, so opsins such as ChR2 did not evolve for this purpose. Nevertheless, ChR2 can still be excited by two-photons at approximately twice the wavelength (~470 nm × 2) of its single-photon activation band. In experimental settings, femtosecond pulsed lasers in the near-infrared range (900 to 1100 nm) are used to drive opsins through two-photon absorption. Because the probability of two-photon absorption scales with the square of photon density, excitation is confined to a tiny focal volume deep within brain tissue. By appropriately shaping the size and geometry of this focal region, it becomes possible to selectively stimulate specific neuronal populations. Techniques that achieve such focal shaping include spiral scanning (tracing the cell perimeter in a spiral to stimulate the entire membrane evenly), temporal focusing, and computer-generated holography (CGH) (**Fig. 1F**). The primary purpose of applying these stimulation methods in vivo to animal brains is to investigate the causal effects of specific stimulation on distinct cell populations and its impact on behavior (**Fig. 1G**). As no other method exists to achieve this except two-photon optogenetics, two-photon optogenetic stimulation remains a challenging yet promising technique for future development.

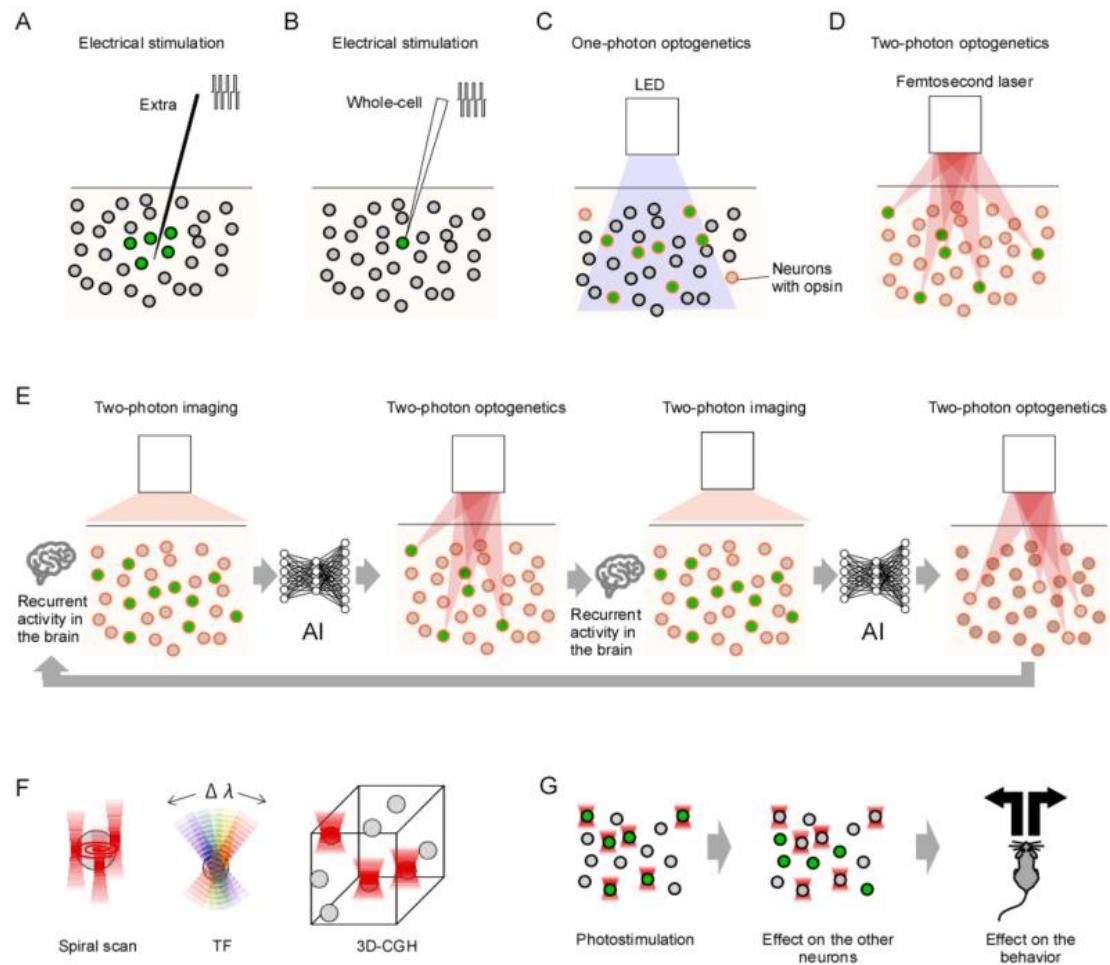

**Figure 1. Mechanisms and purposes of two-photon optogenetic stimulation of neurons**

**A.** Extracellular electrical stimulation with electrodes activates cells near the stimulating site, and if axons run nearby they may also be recruited.

**B.** When current is injected through whole-cell patch clamp, stimulation can be confined to a single neuron, although the number of neurons that can be patched simultaneously is limited to only a few.

**C.** One-photon optogenetic stimulation, based on opsin expression and light delivery, can spatially restrict activation to opsin-expressing cells, but it cannot be confined to a precisely targeted single cell.

**D.** Two-photon optogenetic stimulation overcomes this limitation: owing to the nonlinearity of two-photon absorption, opsin expression does not need to be spatially restricted, and arbitrary stimulation patterns can be generated.

**E.** Two-photon calcium imaging can be used to monitor brain activity and feed it into AI, which in turn guides two-photon optogenetic stimulation. Iterating this closed-loop interaction between brain and AI may open new directions for brain–machine interfaces.

**F.** Three major strategies for two-photon optogenetic stimulation have been developed—spiral scanning, temporal focusing (TF), and three-dimensional computer-generated holography (3D-CGH).

**G.** By selectively stimulating single cells or ensembles and monitoring the effects on non-targeted cells and behavior, researchers can directly probe causal relationships between neural circuits and behavior at the ensemble level.

## 1-2. Differences Between Two-Photon Optogenetic Stimulation and Two-Photon Fluorescence Imaging

Two-photon calcium imaging has become a widely used approach for monitoring neuronal activity during behavior (Hira et al. 2013). The most commonly used fluorophore for this purpose is G-CaMP or GCaMP (Nakai, Ohkura, and Imoto 2001; Zhang et al. 2023), a GFP variant engineered to respond to changes in intracellular $Ca^{2+}$ concentration. GFP and GCaMP emit fluorescence within 5 ns after excitation. This ~5 ns fluorescence lifetime is shorter than the 12.5 ns pulse interval of the 80 MHz ultrafast lasers typically used for two-photon excitation. As a result, regardless of whether the previous pulse has already excited the fluorophore, the molecule is ready to be re-excited by the next pulse. This property makes 80 MHz pulse trains highly efficient for fluorescence imaging.

By contrast, in two-photon optogenetic stimulation, the excited molecule is an opsin such as ChR2. Upon excitation, the ion channel undergoes a conformational change that opens an ionic pore and then closes again, with opening and closing kinetics on the order of 10 ms. Compared with the ~5 ns fluorescence lifetime, this corresponds to a difference of about six orders of magnitude. During this 10 ms period, an 80 MHz laser delivers roughly one million pulses. If the ion channel is opened by the first pulse, the subsequent pulses provide no additional effect. Therefore, the laser pulses must be delivered at a much lower repetition rate. This vast disparity in time scales underlies the fundamental distinction in the choice of optimal light sources for imaging versus stimulation.

The differences between two-photon fluorescence imaging and two-photon optogenetic stimulation are not only temporal but also spatial. Imaging biological specimens typically requires submicron, near-diffraction-limited resolution to visualize intracellular structures and the fine morphology of dendritic spines and axons. In contrast, when two-photon optogenetic stimulation is applied in vivo, the target is often a single cell. In such cases, a resolution of about 10 μm, corresponding to the size of a soma, is sufficient. If the stimulation spot is ~1 μm in diameter, it cannot cover an entire soma and must be scanned across the membrane. Conversely, if the lateral focus is broadened to ~10 μm by spatial focusing alone, the axial confinement of two-photon excitation is severely compromised. Thus, precise three-dimensional control of the excitation volume is required, creating fundamental optical differences between imaging and stimulation.

**Figure 2A** schematically shows the optical path of a typical two-photon microscope for imaging. Images are obtained by scanning the focus in 3-D: an electro tunable lens (ETL) placed at a pupil (Fourier) plane controls axial position, and XY scanners (galvo or resonant) control lateral scanning. With this configuration, three-dimensional two-photon optogenetic stimulation is also possible, but per-cell dwell times of ≥10 ms, together with additional focus travel times of ~10 ms between cells, are often too slow. While this may be sufficient for certain experiments, it is clearly inadequate when the goal is to activate dozens of assemblies simultaneously or to investigate spike timing–dependent plasticity (STDP), where millisecond-scale temporal precision is essential (Caporale and Dan 2008).

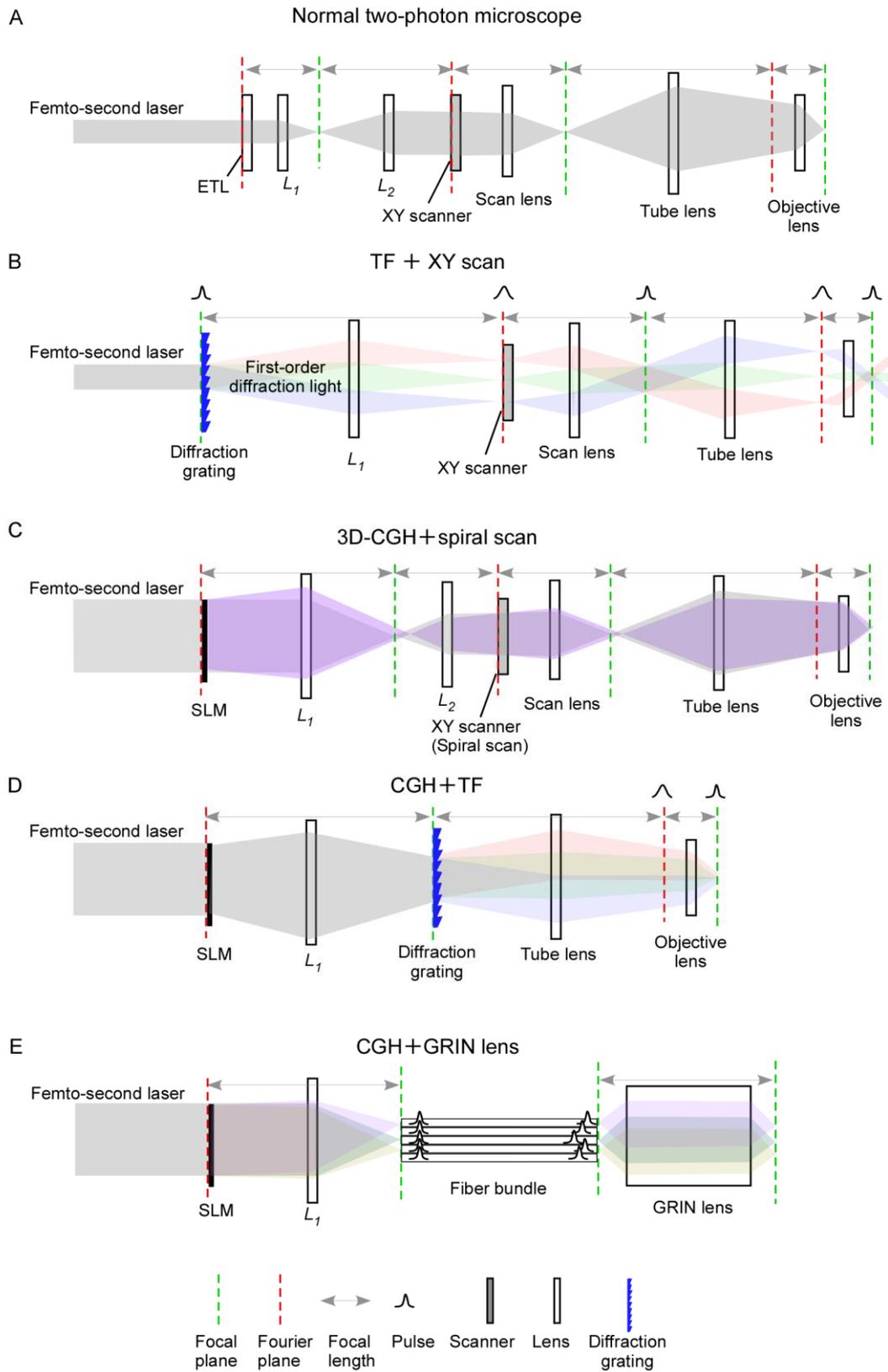

**Figure 2. Methods of two-photon optogenetics**

**A.** A standard two-photon microscope primarily used for imaging. A variable-focus lens at the pupil (Fourier) plane controls the axial position, while XY scanners (galvo or resonant) move the lateral focus. Although this architecture can in principle be used for three-dimensional two-photon optogenetic stimulation, ChR2 activation typically saturates under standard conditions.

**B.** Two-photon stimulation by temporal focusing. A grating conjugate to the focal plane disperses wavelengths at different angles, which recombine only at the focal plane to restore a short pulse. As a result, two-photon excitation occurs only near the focal plane, thereby improving axial resolution.

**C.** Two-photon stimulation using three-dimensional holography. An SLM placed at the pupil (Fourier) plane generates stimulation spots on the focal plane (gray) and, by adding spherical phase, at different depths (purple). Spiral scanning with galvo mirrors expands each holographic spot to the size of a soma, efficiently activating opsins across the membrane.

**D.** Two-photon stimulation combining holography and temporal focusing. To stimulate without scanners, the holographic focal volume itself must approximate the size of a soma. However, if the lateral FWHM is extended to ~10 μm by spatial focusing alone, axial resolution collapses. By placing a grating conjugate to the focal plane after the SLM, spectral dispersion ensures that temporal focusing occurs only at the focal plane, thereby restoring axial resolution. However, only the plane conjugate to the grating undergoes temporal focusing, limiting stimulation to two dimensions per exposure.

**E.** Transferring a 2-D hologram through a fiber bundle and GRIN lens to the brain. ~2m bundles exhibit picoseconds-scale inter-fiber delay variations, suppressing interference and improving two-photon specificity.

## 2. Development of Two-Photon Optogenetic Stimulation

This section outlines several approaches to two-photon optogenetic stimulation. **Figure 1F** illustrates three fundamental techniques: spiral scanning, temporal focusing (TF), and three-dimensional computer-generated holography (3D-CGH). Many studies employ one of these methods or combinations thereof. In doing so, they probe causal relationships between neural circuits and behavior by examining how stimulation affects non-targeted cells and behavioral outputs (**Fig. 1G**).

In the following, we describe the principles and strategies underlying each technique (and their combinations), followed by brief summaries of representative methods and findings from individual studies. The presentation is arranged roughly chronologically, beginning with early demonstrations and progressing toward increasingly sophisticated combinations designed to meet the distinct spatiotemporal requirements of two-photon excitation for imaging versus stimulation.

### 2-1. Spiral Scanning

Six years after the first report of one-photon optogenetic stimulation (Nagel et al. 2003), the first study on two-photon optogenetic stimulation was published in 2009 (Rickgauer and Tank 2009). In cultured neurons, they demonstrated that ChR2 could be driven by two-photon excitation to elicit action potentials. The stimulation employed a Ti:sapphire laser at ~920 nm and 80 MHz, parameters typical for two-photon imaging at the time. As emphasized in the title of their paper, ChR2 activation rapidly saturated within the focal region: whereas fluorophores have fluorescence lifetimes of ≤10 ns, ChR2 channel kinetics are on the order of ~10 ms—a difference of approximately six orders of magnitude. Consequently, immediately after two-photon stimulation, most ChR2 molecules within the focal spot are in the open state, and further illumination there becomes effectively meaningless. At the same time, even regions adjacent to the focal volume with slightly lower photon density can still activate ChR2: when the per-pulse opening probability is small but nonzero, the enormous number of pulses eventually induces channel opening. Thus, the spatial resolution of two-photon optogenetic stimulation is constrained by mechanisms distinct from those governing fluorophore excitation. To optimize both specificity and speed, Rickgauer and Tank (Rickgauer and Tank 2009) reduced the effective NA of the objective to axially elongate the PSF, which was then spiral scanned along the cell outline (**Fig. 1F**). This strategy successfully induced single-cell action potentials.

This pioneering study appears to have set two directions for subsequent work on two-photon opsin excitation: (i) tailoring the spatial region where photon density is sufficient for two-photon activation to match cellular dimensions, and (ii) exploring the use of low-repetition-rate lasers to mitigate saturation. The first principle has remained central as CGH approaches have been developed for two-photon microscopes, while the second contributed to the trend toward ~1 MHz high-power lasers.

Among later reports employing spiral scanning to achieve single-cell-resolution stimulation (Packer et al. 2012; Prakash et al. 2012), one notable in vivo application was reported by Chettih and Harvey (Chettih and Harvey 2019) for mapping cortical circuits. They repeatedly applied spiral scanning to layer 2/3 neurons in mouse visual cortex (V1) using C1V1 (Yizhar et al. 2011) or ChrimsonR (Klapoetke et al. 2014), restricting opsin expression to perisomatic membranes with a KV2.1 motif (Baker et al. 2016). Stimulation was delivered through a Nikon 16×/0.8 NA objective with a 1070 nm Fidelity 2 laser (a ~2 W, 80 MHz, 100 fs fiber source), while imaging was performed at 920 nm. Cells were stimulated at ~1 cell/s using 12–15 µm spiral scans lasting 32 ms, with an average power of ~52.7 mW; the scan diameter, slightly larger than the soma, accommodated brain motion. Opsin expression was kept sparse, and off-target effects on neighboring neurons were carefully assessed. Stimulation was paired with visual inputs. The authors estimated that ~2500 trials per pair would be required to detect reliable causal effects. Although detailed results fall outside the scope of this technological overview, one noteworthy finding was that the net effects of optical stimulation were predominantly inhibitory—a theme that has since been echoed in many subsequent experiments.

## 2-2. Temporal Focusing

Temporal focusing (TF) improves the axial resolution of two-photon excitation by placing a diffraction grating or diffuser conjugate to the focal plane (Oron, Tal, and Silberberg 2005). This configuration stretches pulses away from the focal plane and recompresses them only at the focal plane (Andrasfalvy et al. 2010; Papagiakoumou, Ronzitti, and Emiliani 2020). With a diffraction grating, light is spectrally dispersed at planes above and below the focus and recombines only at the focal plane (**Fig. 1F**). Although gratings are commonly used, diffusers that impose random phase shifts can similarly elongate pulses away from the focal plane, and the two approaches can be combined (Mardinly et al. 2018; Papagiakoumou, Ronzitti, and Emiliani 2020).

For neuronal applications, the ideal case is when the focal volume in which peak laser power reaches the two-photon activation threshold matches the size of a single neuron. In that situation, membrane-expressed opsins can be activated without scanning, allowing near-instantaneous activation or inhibition. By expanding the beam laterally at the grating, one can create a disc of ~10–20 µm diameter at the objective's focal plane. If ~10 µm lateral excitation were attempted using spatial focusing alone, axial resolution would be severely degraded. TF mitigates this problem, enabling soma-sized excitation spots.

Rickgauer et al. (Rickgauer, Deisseroth, and Tank 2014) efficiently implemented two-photon optogenetic stimulation with TF as an alternative to spiral scanning (**Fig. 2B**), successfully activating place cell ensembles in hippocampal CA1. The excitation spots were 10–15 μm in diameter with an axial FWHM of ~6 μm. The stimulation laser was a 1064 nm, 5 W source operating at 80 MHz with ~200 fs pulses, while imaging used 920 nm excitation with GCaMP3. Because CA1 has a laminar organization with densely packed neurons in a quasi-two-dimensional sheet, and principal cells lack strong recurrent connectivity, the interpretation of stimulation effects is in some respects more straightforward than in highly recurrent three-dimensional neocortical circuits.

**2-3. Holography + Spiral Scanning**

An LCoS-SLM (liquid crystal on silicon spatial light modulator; hereafter SLM) modulates the phase of the laser wavefront in a position-dependent manner by adjusting the local orientation of liquid crystal molecules to impose phase delays. Typical refresh rates are on the order of tens of milliseconds, while the fastest devices can reach millisecond scales (e.g., Meadowlark products). When placed conjugate to the objective pupil (Fourier plane), the SLM generates holograms at the focal plane (Nikolenko et al. 2008). For ensemble stimulation, the goal is to achieve simultaneous two-photon excitation of specific groups of neurons using holographic patterns. The ability to stimulate multiple neurons at once is a unique advantage of the SLM approach. Methods based on spiral scanning or TF alone cannot accomplish this. Moreover, three-dimensional holograms can be readily obtained by adding spherical phase to a two-dimensional pattern (3D-CGH). If one can design patterns that provide sufficient power for two-photon activation of target neurons without affecting others, and switch them rapidly, it becomes possible to test how different ensembles influence local circuit activity and behavior. The strategy introduced here is to generate a holographic spot at each target neuron's location via the SLM, and then spiral scan each spot to match soma size, thereby achieving effective two-photon optogenetic stimulation (**Fig. 2C**). This remains the most widely used method today, with numerous applications described below. Over time, the adoption of lower-repetition-rate lasers and techniques for peri-somatic opsin localization have become standard.

As a representative in vivo application, Packer et al. (Packer et al. 2015) targeted layer 2/3 pyramidal cells in mouse primary somatosensory (barrel) cortex expressing C1V1. The SLM (7.68 × 7.68 mm active area, 512 × 512 pixels; optimized for 1064 nm; Meadowlark/Boulder Nonlinear Systems) was placed conjugate to the pupil. Stimulation was delivered at 1064 nm (5

W, 350–400 fs; Fianium Ltd.) or 1055 nm (2.3 W, 100 fs; Coherent). A Gerchberg–Saxton (GS) algorithm (Gerchberg 1972) was used to optimize the phase pattern, with pre-compensation to increase peripheral brightness relative to the center, a common practice. Imaging employed GCaMP6s with 920 nm excitation, using a Nikon 16×/0.8 NA water-immersion objective. Spiral scanning (20 μm diameter, 20 ms) enabled simultaneous stimulation of ~10 cells, with a mean spike latency of ~17 ms.

Yang et al. (Yang et al. 2018) performed three-dimensional multi-neuron stimulation in layer 2/3 of mouse V1 using the opsin C1V1. Stimulation was delivered with a Spirit 1040-8 laser (1040 nm, 400 fs, 40 μJ, 0.2–1 MHz; Spectra Physics), while imaging was conducted at 940 nm. They employed an HSP512-1064 SLM (7.68 × 7.68 mm² active area, 512 × 512 pixels; Meadowlark) in combination with an Olympus 25×/1.05 NA XLPlan water-immersion objective. Approximately 50 cells distributed across ~150 μm depth were targeted, and 78% of them were activated with ~300 mW for 95 ms, although non-targeted cells within ~20 μm also showed notable activation. In addition, they expressed C1V1 in SOM interneurons via Cre drivers and selectively stimulated these cells.

Carrillo-Reid et al. (Carrillo-Reid et al. 2019) used two-photon holography in mouse V1 to selectively activate learned functional ensembles and modulate visual behavior. They expressed the opsin C1V1 in layer 2/3 pyramidal neurons, and stimulation was delivered at 1040 nm with a 1 MHz laser. Imaging was performed at 940 nm with GCaMP6s using the same Olympus 25×/1.05 NA objective. Stimulation patterns were generated by combining an SLM with spiral scanning, using 12 μm-diameter spirals at 20 Hz with ~5 mW per cell. During the "Go" window of the behavioral task, illumination lasted 2 s. Remarkably, activating only two neurons from a learned ensemble was sufficient to recruit the rest of the ensemble and enhance visual discrimination performance, whereas perturbing the ensemble by stimulating unrelated neurons degraded performance.

In another study, Marshel et al. (Marshel et al. 2019) employed two-photon holography to reproduce visual percepts by stimulating excitatory ensembles in mouse V1. They introduced ChRmine, an opsin generating exceptionally high photocurrent. Stimulation was delivered with a Monaco 1035-80-60 laser (1035 nm, 10 MHz, 80 W, 300 fs). Imaging was performed at 940 nm. A custom MacroSLM (1536 × 1536 pixels, 30.7 × 30.7 mm active area, 20 μm pitch) was used in conjunction with spiral scanning, in which 25 μm-diameter spirals were applied for 4 ms and repeated 12 times. By replaying spatial patterns of naturally evoked activity, they were able to modulate the animals' behavioral accuracy, with simultaneous targets ranging from several dozen neurons up to approximately 100 cells.

Dalgleish et al. (Dalgleish et al. 2020) expressed C1V1-KV2.1 in layer 2/3 pyramidal cells of the S1 barrel cortex and applied two-photon stimulation using an SLM combined with spiral scanning. Stimulation was delivered with an Amplitude Satsuma laser operating at 1030 nm, 2 MHz, and 20 W. The SLM was a Meadowlark device with 512 × 512 pixels and a 7.68 × 7.68 mm active area. Spiral scans were 10 μm in diameter, with ~6 mW per cell, and each trial involved ~5 ms of stimulation. The resulting spatial resolution was ~5 μm laterally and ~20 μm axially. By varying the number of neurons stimulated per trial, the authors estimated that synchronous activation of ~14 neurons was sufficient to overcome network inhibition and improve detection performance.

Fişek et al. (Fişek et al. 2023) investigated feedback from higher visual area LM to V1 by stimulating neurons in LM. They expressed C1V1-KV2.1 and used an Amplitude Satsuma HP2 laser operating at 1 MHz, 30 W, 40 μJ, and 400 fs. Two microscope configurations were employed: one with a Nikon 16×/0.8 NA objective and an SLM with 512 × 512 pixels and a 7.68 × 7.68 mm active area (BNS/Meadowlark), and another with a Thorlabs 10×/0.5 NA objective and a larger SLM with 1920 × 1152 pixels and a 17.6 × 10.7 mm active area (Meadowlark). These provided fields of view of 1215 μm and 1920 μm, respectively. Stimulation was applied with SLM-guided spiral scanning, delivering ~12 mW per cell with 16 μm-diameter spirals, using 10 ms pulses at 20 Hz repeated 10 times for a total duration of 500 ms. Functionally, they found that paired LM and V1 neurons with overlapping receptive fields exhibited suppression in V1, while pairs with slightly different receptive fields showed excitation, attributable to apical dendritic $Ca^{2+}$ spikes. The top-down suppression observed in the case of shared receptive fields is consistent with predictive coding, whereas the excitation seen with non-identical receptive fields may support context-dependent cooperation. From a technical perspective, their design leveraged ~1 mm-scale FOV to probe inter-areal feedback by combining two-photon stimulation with dendritic imaging.

Russell et al. (Russell et al. 2024) assessed the sensitivity of two-photon stimulation in mouse V1 during a contrast detection task. They expressed C1V1-KV2.1 and used a Satsuma laser with a 512 × 512 SLM, following the setup described in Fişek et al. (2023), but explicitly implemented the Bruker NeuraLight 3D system. A three-dimensional holographic pattern combined with spiral scanning delivered 10 μm-diameter, 20 ms stimulations once every 50 ms during the behavioral task. When visual contrast was low and the pupils were dilated, indicating engagement, optical stimulation improved detection. The authors concluded that cortical responses to stimulation depend strongly on behavioral and affective state as well as sensory context.

Wu, Chen et al. (Wu et al. 2025) targeted the motor cortex using peri-somatically localized ChRmine-KV2.1 delivered with a commercial Bruker two-photon system. Stimulation was performed with a Monaco 1035-40-40 laser (1 MHz, 1035 nm, 40 W, 350 fs), and holograms were generated with a 512 × 512 SLM operating at an update rate of ~600 Hz. Spiral scanning was applied with 10 μm-diameter rotations at ~7 mW per cell. Using the NeuraLight 3D system—also employed in Piantadosi et al. (2024), Russell et al. (2024), and Vinograd et al. (2024) (Piantadosi et al. 2024; Russell et al. 2024; Vinograd et al. 2024)—they demonstrated that stimulating ~20 early task-related M1 neurons was sufficient to initiate movement, whereas stimulating ~10 early neurons or late-phase neurons was not. Movement initiation depended on the preparatory state, and stimulation of early neurons recruited additional non-stimulated, movement-related neurons, consistent with a pattern-completion mechanism.

## 2-4. Holography + Temporal Focusing

In earlier approaches, TF and holography were applied separately, with a galvo mirror placed downstream of either the grating or the SLM to direct patterned light to the desired location. Here we discuss a key advance: directly setting the lateral size of the hologram to match the soma scale, while using temporal focusing to restore axial confinement. This approach enables stimulation without spiral scanning (**Fig. 2D**).

Papagiakoumou et al. (Papagiakoumou et al. 2010; Papagiakoumou et al. 2009; Papagiakoumou et al. 2008) first implemented this strategy by sending light to an SLM to generate a phase pattern, projecting that pattern onto a grating, and then achieving temporal focusing at the objective focal plane. Because the SLM and the grating are positioned at Fourier-related planes, simply placing the grating first would cause spectrally separated spots to appear on the SLM, making it apparently impossible to construct a general pattern through wavelength-dependent phase. For this reason, the sequence SLM followed by grating is required.

Chen et al. (Chen et al. 2019) demonstrated sub-millisecond jitter and tens-of-hertz spiking in layer 2/3 of V1 using a combination of temporal focusing and holography with ReaChR, CoChR, and ChrimsonR. Stimulation was provided by an Amplitude Satsuma HP laser (1030 nm, 250 fs, 20 μJ; 10 W, 500 kHz), and the SLM was a Hamamatsu X10468-07 (792 × 600 pixels, 20 μm pitch, rise time 10 ms, fall time 80 ms).

A limitation of the SLM-to-grating configuration is that two-photon excitation is effectively restricted to the two-dimensional pattern formed on the grating plane. Even if the SLM phase pattern is designed as a three-dimensional hologram, spectral dispersion occurs only at the grating-conjugate plane, and temporal focusing is confined to that plane. One possible approach to realizing a three-dimensional hologram is to add spherical phase terms to each two-dimensional layer and sum them. However, outside the grating-conjugate plane, differences in optical path length across wavelengths prevent pulse recompression, making temporal focusing ineffective. As a result, spatial resolution for stimulation at planes other than the conjugate plane deteriorates substantially. The next section describes 3D-SHOT, which overcomes this limitation.

## 2-5. 3D-SHOT (Three-Dimensional Scanless Holographic Optogenetics with Temporal Focusing)

As noted, 3D-SHOT combines temporal focusing with SLM-based three-dimensional holography to achieve scanless stimulation of ensembles distributed in three dimensions. In standard holography, all wavelengths follow the same optical path. In contrast, 3D-SHOT introduces a grating that first disperses the spectrum, allowing the SLM to impose wavelength-dependent phase shifts and thereby enabling temporal focusing at each target spot. This approach was proposed and demonstrated in two studies from the Adesnik lab (Mardinly et al. 2018; Pégard et al. 2017). The 2018 implementation that introduced a rotating diffuser ("3D-SHOT 2.0") is now widely employed, while the 2017 study laid out the conceptual framework.

Pégard et al. (Pégard et al. 2017) introduced the theoretical framework and demonstrated the fusion of three-dimensional holography with temporal focusing (**Fig. 3A**). They used ChrimsonR and Chronos as opsins, restricting their expression to peri-somatic membranes with a KV2.1 motif. A modified multilevel Gerchberg–Saxton algorithm was developed, incorporating analytic expressions for the target phase.

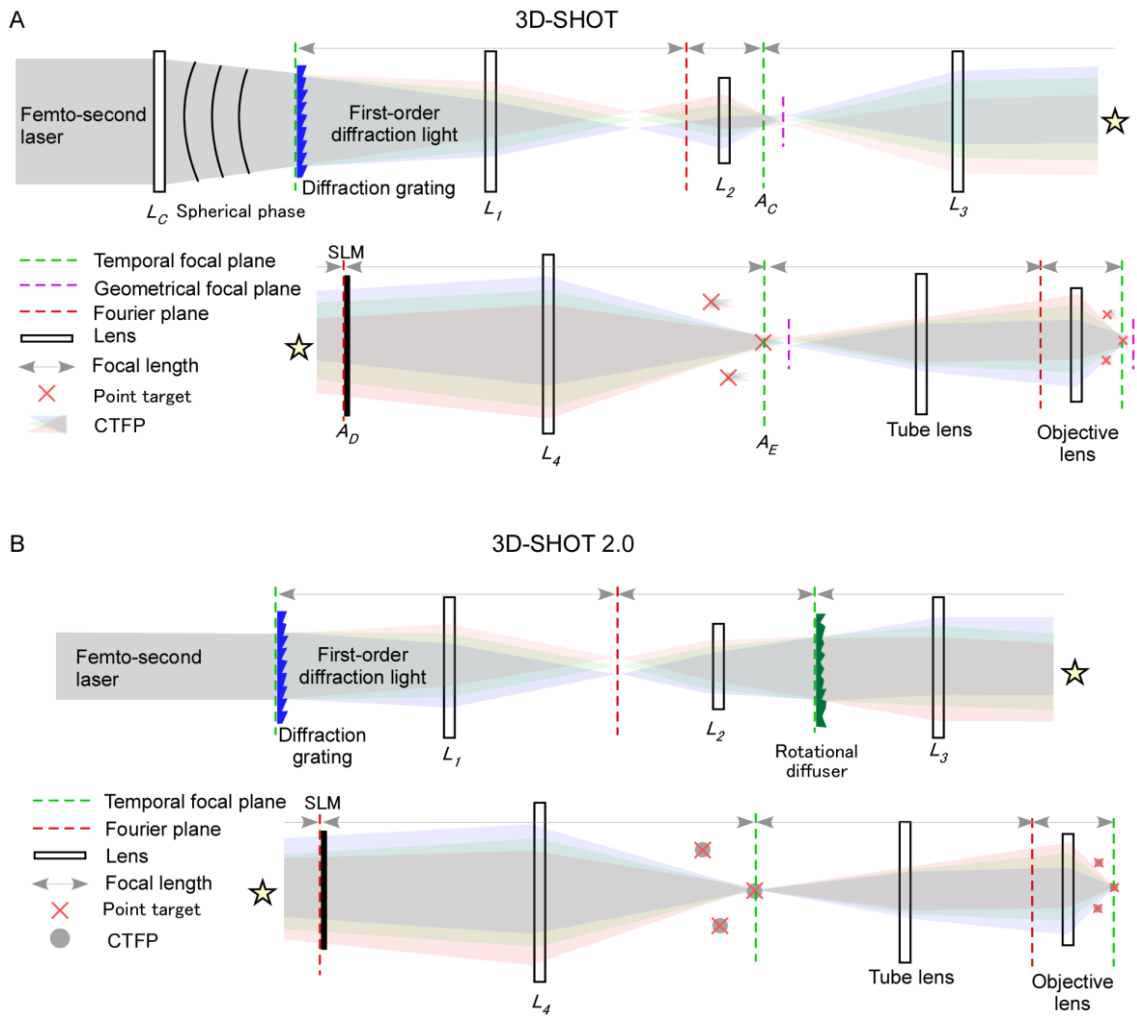

**Figure 3. Mechanism of 3D-SHOT.**

**A.** 3D-SHOT optical path. A grating is placed before the SLM. A lens Lc adds spherical phase to the light incident on the grating. By intentionally separating the temporal-focus plane and the geometric focus plane, 3D-CGH is reconciled with TF. The custom temporal-focus pattern (CTFP) formed by the grating + Lc is copied by the SLM to arbitrary 3-D positions. Because different wavelengths illuminate different SLM pixels, the SLM can adjust phase per wavelength, equalizing path length and realizing TF at each 3-D target.

**B.** 3D-SHOT 2.0. A rotating diffuser substitutes for Lc. The diffuser slightly spreads each wavelength at the SLM, enabling both spatial CGH and per-wavelength phase control to equalize path lengths for TF, while speckle is temporally randomized.

In 3D-SHOT, the grating together with the preceding lens (Lc) first generates a disk-shaped temporal focusing pattern, referred to as the custom temporal focus pattern (CTFP). The SLM then replicates this CTFP at arbitrary three-dimensional coordinates as a point-cloud hologram. Importantly, the SLM does not sculpt the focal spot itself; it merely places copies of the CTFP at designated 3D positions. The incident light is converted into a spherical wave by Lc before reaching the diffraction grating. The plane of the grating corresponds to the temporal focus plane (green). After the grating, the light is relayed by a 4f system consisting of L1 and L2. The Fourier plane (red) and the temporal focus plane lies between L1 and L2, shifted closer to L2 relative to the geometrical focal plane. This Fourier plane is conjugate to the SLM. The temporal focus plane, labeled Ac, is offset from the subsequent geometric focal plane—an intentional displacement engineered by Lc—which generates a conical region of high photon density (the CTFP). Note that the photon density may split into two peaks if the distance between these planes is large. The SLM is placed, via L3, at the Fourier plane of the temporal focus plane, where a phase pattern is applied. Here, the phase pattern is computed to replicate the CTFP in three dimensions. The SLM thus produces a point-cloud hologram, but unlike conventional holography, different wavelengths strike different SLM pixels, allowing the imposition of wavelength-specific phases. This enables not only ordinary 3D spatial focusing according to the holographic design but also temporal focusing at each target point by equalizing the optical path length across wavelengths. This feature distinguishes 3D-SHOT from simple 3D-CGH. Although the calculations are intricate, details are provided in the Supplementary Information of the paper. After L4, the replicated CTFPs distributed in 3D are transferred through the tube lens and objective to the neural population, where each site undergoes temporal focusing. The key innovation of 3D-SHOT lies in the arrangement of the grating and Lc, which allows the SLM to be used for combined spatiotemporal focusing.

The next step is an improved version, often referred to as 3D-SHOT 2.0 (**Fig. 3B**). Mardinly et al. (Mardinly et al. 2018) successfully achieved efficient, bidirectional control of three-dimensionally distributed neuronal ensembles in the mouse cortex. The opsins used were ChroME (an excitatory opsin) or eGtACR1 (an inhibitory, Cl$^-$-conducting opsin), each fused to a Kv2.1 trafficking sequence. The excitation sources were either a Coherent Monaco laser (1040 nm, 2 MHz, 40 W) or an Amplitude Satsuma laser (1040 nm, 2 MHz, 20 W). The SLM was a Meadowlark 512 L. A rotating diffuser ($\theta = 0.5°$, #47-989, Edmund Optics) was mounted on a motor with a DVD spindle and served as a replacement for lens Lc.

In 3D-SHOT 2.0, lens Lc is omitted, and the incident beam is directed straight onto the diffraction grating, which is conjugate to the focal plane. After lens L1, the spectrally dispersed light converges at a plane conjugate to the Fourier plane. Downstream of L2, a rotating diffuser is positioned at a plane conjugate to the grating. Here, the diffuser introduces random phase shifts, which mitigate frequency-dependent focusing when projected onto the SLM, placed at the Fourier plane after L3. This reproduces the functionality originally provided by Lc in the first 3D-SHOT implementation. Without the diffuser, the dispersed spectrum would be sharply aligned across the SLM, as at a pupil plane between L1 and L2. With the diffuser, however, light of different frequencies spreads laterally and overlaps on the SLM, as shown in the schematic. Rotation of the diffuser temporally randomizes speckle, reducing unwanted stimulation outside the designed pattern. In addition, because the diffuser broadens the local pulse duration, it contributes to temporal focusing. Thus, without sacrificing temporal focusing by the grating, 3D-SHOT 2.0 introduces slight spectral spread that allows wavelength-specific phase control at the SLM while retaining ordinary spatial holographic control.

Pattern generation was implemented using either the Gerchberg–Saxton (GS) algorithm or the NOVO-CGH algorithm (Zhang et al. 2017). With sub-millisecond temporal precision (jitter < 1 ms), they were able to simultaneously stimulate up to 50 neurons in 3D, synthesizing arbitrary firing patterns within a volume of $550 \times 550 \times 100$ μm.

Below, we introduce three studies that have employed 3D-SHOT. In the study by Sridharan et al. (Sridharan et al. 2022), the opsin ChroME was engineered to generate several new variants, including ChroME2s. Although the photocurrents of ChroME2s are weaker than those of ChRmine, the faster on/off kinetics enable reliable high-frequency stimulation. Using 3D-SHOT, they succeeded in simultaneously stimulating a total of 631 cells per second in the mouse visual cortex, divided into multiple groups. The excitation sources were a Satsuma HP2 (1030 nm, 2 MHz, 350 fs, Amplitude) or a Monaco 1035-80-60 (1040 nm, 1 MHz, 300 fs, Coherent), providing up to 60 W of power. A blazed diffraction grating (600 l/mm, 1000 nm blaze; Edmund Optics 49-570 or Newport 33010FL01-520R) was employed, together with an HSP1920 SLM (192 × 1152 pixels; Meadowlark Optics). They adopted the 3D-SHOT 2.0 configuration with a rotating diffuser, and the SLM phase patterns were computed using the Gerchberg–Saxton (GS) algorithm. ChroME2s exhibited large instantaneous photocurrents and fast response kinetics, making it particularly effective for high-temporal-precision applications. However, in terms of overall sensitivity to two-photon optogenetic stimulation, ChRmine still outperformed ChroME2s.

Oldenburg et al. (Oldenburg et al. 2024) used 3D-SHOT to precisely manipulate excitatory populations in layer 2/3 of mouse V1, thereby dissecting recurrent circuit motifs. The opsins were ChroME or ChroME2s. The laser source was a Monaco 40 (1040 nm, 2 MHz), paired with a

blazed diffraction grating (Newport R5000626767-19311) and an HSP1920-1064-HSP8-HB SLM (1920 × 1152 pixels; Meadowlark Optics). The implementation followed the rotating diffuser approach described by Mardinly et al. (Mardinly et al. 2018). Per-cell powers ranged from 12.5 to 100 mW. The achieved axial resolution was ~10–15 μm FWHM, remaining nearly invariant across ~80 μm of depth. Functionally, ensembles with similar visual tuning but extended spatial distributions tended to recruit additional nearby activity, whereas compact ensembles induced strong local inhibition, demonstrating an interplay between physical arrangement and feature preference.

Bounds & Adesnik (Bounds and Adesnik 2024) examined how 3D-SHOT–mediated stimulation of V1 ensembles during a visual detection task modulates both surrounding activity and behavior. The opsin was ChroME2s without soma restriction. The optical setup included a Monaco 1035-80-40 laser (1035 nm, 2 MHz, 276 fs), an HSP1920-1064-HSP8-HB SLM (1920 × 1152 pixels; Meadowlark Optics), and a blazed diffraction grating (Newport 33010FL01-520R). The effective resolution was ~9.7 μm laterally and ~20 μm axially (FWHM). They simultaneously stimulated 7–40 neurons. Strikingly, targeting the most visually responsive cells alone did not yield a special behavioral advantage. Instead, the behavioral outcome depended on how much additional local activity the stimulation recruited, underscoring the role of ensemble context within cortical networks.

### 2-6. 3D-SHOT Combined with Large FOV Two-Photon Microscopy

3D-SHOT is considered one of the definitive methods for fast and flexible three-dimensional stimulation of neuronal ensembles. However, one of the principal limitations on the number of neurons that can be stimulated simultaneously arises from the field of view (FOV) of the objective lens. A natural way to overcome this constraint is to employ large FOV two-photon microscopes. Several such systems have been developed, including Diesel2p (Hira, Imamura, et al. 2025; Hira, Townsend, et al. 2025; Yu et al. 2021), 2p-RAM (Sofroniew et al. 2016), and Fashio-2PM (Ota et al. 2021). Here, we introduce a study that applied 3D-SHOT to the 2p-RAM platform.

Abdeladim et al. (Abdeladim et al. 2023) integrated the 3D-SHOT optical path into a commercial large FOV two-photon microscope (2p-RAM), which provides an imaging FOV exceeding 5 mm. This implementation enabled three-dimensional holographic photostimulation within a ~1 mm subregion of the field. The opsins were *ChroME* or *ChroME2s*. The experimental animals included Vglut1-Cre × Ai203 double-transgenic mice (Bounds et al. 2023), as well as triple-transgenic mice (EMX1-Cre, CaMK2-tTA, tetO-GCaMP6s) into which Cre-dependent *ChroME2s* was delivered via AAV. Two-photon stimulation was performed using a fiber laser

(Aeropulse 50, NKT Photonics; 1030 nm, ~1 MHz, 50 W, <500 fs), and GCaMP7s imaging was carried out at 920 nm. A 1920 × 1152 SLM (Meadowlark Optics) was used. Because the 2p-RAM system employs a vertically mounted rotating breadboard, both stimulation and imaging beams were dynamically aligned via a custom "gantry periscope." The optical path followed the standard 3D-SHOT configuration: diffraction grating → rotating diffuser → SLM → objective.

The achieved spatial resolution was ~32 μm laterally and ~9 μm axially (FWHM), slightly lower than in other 3D-SHOT implementations (20 μm in Mardinly et al. (Mardinly et al. 2018); 28 μm in Pégard et al. (Pégard et al. 2017)), likely due to the relatively low NA (0.6) of the objective. The effective stimulation range was ~1 × 1 mm, which represents only a fraction of the total 2p-RAM FOV. In their experiments, they imaged a ~3 mm-wide region encompassing primary and multiple higher visual areas while stimulating ensembles in the higher visual area LM. Although some distant neurons showed increased activity, the net effect was predominantly suppressive both within LM and in distal regions. This represents the first study to causally examine inter-areal propagation of activity at the level of neuronal ensembles.

As the work remains at the preprint stage, several technical limitations are notable. The full FOV of 2p-RAM was not exploited, and the stimulation region covered only a subset of the available imaging field. In principle, even with a limited SLM aperture, galvanometric scanning could allow the stimulation field to be repositioned anywhere within the imaging FOV, but this has not yet been achieved. Moreover, 2p-RAM suffers from substantial field curvature, with focal shifts of ~150 μm at the edges of the imaging field. This may pose a practical challenge for repositioning the stimulation plane across large distances without physically moving the animal.

## 3. Summary and Comparison of Two-Photon Optogenetic Stimulation Methods

In Section 2, we introduced the development of two-photon optogenetic stimulation methods as a trajectory toward the goal of applying 3D-SHOT to large FOV two-photon microscopy. The field progressed from early proof-of-principle studies using spiral scanning, to the adoption of temporal focusing and holography, and eventually to their combination for three-dimensional implementations. In practice, however, the approach most commonly chosen and successfully applied in neuroscience experiments at present is to generate two- or three-dimensional holograms with an SLM and scan them with galvanometric mirrors.

The fundamental distinction between this method and 3D-SHOT lies in whether temporal focusing or galvanometric scanning is used. In terms of stimulation speed, 3D-SHOT is advantageous, as opsins distributed across the membrane can be activated simultaneously, whereas spiral scanning is inherently limited by the millisecond-scale response time of the galvos. By contrast, laser power requirements favor the hologram-plus-spiral approach: by generating diffraction-limited holographic foci and scanning them across a soma, it is possible to achieve activation with lower average power. In 3D-SHOT, however, the light intensity must reach two-photon excitation thresholds over an entire soma-sized (or larger) volume, which demands substantially higher power. As a result, compared to holographic spiral scanning, 3D-SHOT offers lower latency and jitter, but at the cost of requiring high-power lasers and posing an increased risk of tissue heating. The potential for tissue damage due to elevated laser power is a serious concern. Taken together, holographic spiral scanning remains the more practical and widely used approach at present. If future technical advances resolve the power requirement, however, 3D-SHOT may become the more suitable method.

Another important point is that while two-photon optogenetic stimulation represents an optimal tool for manipulating the activity of precisely defined neuronal ensembles, such technically elaborate experiments are not always necessary depending on the experimental aim. For example, in our previous work (Hira et al. 2014), opsins were sparsely expressed in ~5% of neurons in motor cortex. One-photon stimulation was applied during two-photon calcium imaging, while water rewards were delivered either before or after photostimulation. Within only 15 minutes of training, mice increased the spontaneous activity of stimulated neurons when stimulation preceded reward, but decreased their activity when reward preceded stimulation. This bidirectional modulation may represent a fundamental mechanism underlying the association between actions and rewards.

Similarly, in the study by Carrillo-Reid et al., (Carrillo-Reid et al. 2016) opsins were expressed in visual cortical neurons, and a two-photon stimulation laser was scanned across the entire imaging field at specific time points to activate all opsin-expressing neurons within view. Repeated stimulation imprinted a functional ensemble independent of visual input, which maintained its functional connectivity across days. Strikingly, stimulation of a single neuron within this ensemble was sufficient to reactivate the entire ensemble. The authors referred to this process as *pattern completion*, and the reactivation as *recall*.

These findings demonstrate that meaningful manipulations of ensemble activity, and their behavioral consequences, can often be achieved with simpler strategies—such as sparse opsin expression combined with one-photon stimulation, or large FOV two-photon scanning—despite their reduced specificity and flexibility compared to advanced approaches like 3D-SHOT. Thus, experimenters should avoid being overly captivated by technically "fancy" methods, and instead design experiments in a purpose-driven manner aligned with the scientific question at hand.

Although this article has emphasized optical methodologies, it is also worth briefly summarizing what has been learned from two-photon optogenetic stimulation itself. A simple yet striking finding across many studies is that photostimulation of excitatory neurons generally produced net suppressive effects in local circuits. This suppressive influence was observed both in the immediate neighborhood of the stimulated cells and in distant cortical regions. These results clearly indicate that cortical circuits maintain a finely tuned balance between excitation and inhibition, such that additional excitatory drive engages inhibitory interneurons and normalizes the overall spike output to approximately zero net change.

Nevertheless, several studies have reported *pattern completion*, whereby activation of a subset of neurons can trigger activity across an entire functional ensemble. This suggests that while the overall circuit response tends to be suppressive, functionally related subsets of neurons can still be strongly activated. Such observations are consistent with the known anatomical organization of local cortical networks (Yu et al. 2025).

By contrast, the behavioral consequences of stimulation appear to vary substantially with context, and thus require careful re-examination under diverse conditions and brain states before firm conclusions can be drawn. Given the inherent complexity of the brain, the effects of stimulation in specific cognitive tasks or contexts may lead to different interpretations depending on subtle differences in experimental design. These technically demanding experiments therefore require not only cautious interpretation of results but also close evaluation of the methodological assumptions underlying them.

## 4. Related Techniques and Outstanding Issues

In this section, we briefly summarize additional technical advances and challenges in two-photon optogenetic stimulation that were not covered in the previous discussion. In 3D-SHOT, because the disk-shaped pattern at the entrance of the diffraction grating is replicated by the SLM and distributed in three dimensions, the freedom to control spot size and shape under the objective lens is limited. To address this, Accanto et al. (Accanto et al. 2018) introduced a second SLM before the grating, enabling arbitrary control over the input shape. This approach, termed *multiplexed temporally focused CGH (MTF-CGH)*, successfully enhanced the flexibility of optogenetic stimulation in terms of spot size and geometry.

Similarly, Faini et al. (Faini et al. 2023) proposed a strategy called *FLiT*, in which a galvanometric mirror is placed upstream of the SLM to divide the SLM surface into multiple regions. Although updating an SLM phase pattern typically requires several milliseconds, this configuration allows rapid selection of preloaded holograms with the galvo, achieving sub-millisecond updates of stimulation patterns.

Inazawa et al. (Inazawa et al. 2021) employed an echelle grating to introduce optical path differences larger than the pulse width in the direction orthogonal to spectral dispersion. In combination with a DMD that generated multiple line segments, fast and accurate acquisition of optical sectioning was achieved (*time-multiplexed multifocal temporal focusing*). Ishikawa et al. (Ishikawa et al. 2021) extended this by adding an SLM, applying CGH to suppress interference fringes and speckle, a method that should also be applicable to two-photon stimulation.

While galvanometric mirrors require milliseconds to switch stimulation between individual cells, so-called random-access scanning with acousto-optic deflectors (AODs) can reduce this timescale to the microsecond range. This is faster than the intrinsic timescale of neuronal ensemble dynamics, making AOD-based approaches promising for two-photon optogenetic stimulation with diffraction-limited beams. Indeed, AODs are already widely used in fast two-photon imaging (Geiller et al. 2020; Griffiths et al. 2020; Judák et al. 2022; Villette et al. 2019). However, no reports have yet demonstrated large-scale three-dimensional two-photon optogenetic stimulation using AODs. This may reflect fundamental limitations: because AODs create grating-like patterns using ultrasound, different wavelengths become spatially separated, leading to pulse broadening; moreover, the scanning angle is inherently restricted. Nevertheless, the commercial availability of the FEMTO3D Atlas (Femtonics), which supports AOD-based two-photon optogenetics, indicates that the technology has reached sufficient stability for routine use, and it may become more widely adopted in the near future.

We next turn to algorithms for generating computer-generated holography (CGH). In most studies, the phase masks displayed on the SLM are computed using the Gerchberg–Saxton (GS) algorithm (Gerchberg 1972). The GS algorithm iteratively optimizes the SLM phase pattern by alternately applying Fourier and inverse Fourier transforms between the Fourier plane, where the SLM is located, and the focal plane. Since SLMs generally modulate only the phase, the intensity distribution of the incident coherent laser beam directly constrains the reflected intensity distribution, which imposes a boundary condition on the optimization. A common issue is that naïve implementations produce weaker illumination at the periphery of the field relative to the center. To compensate for this, preprocessing steps—such as modifying the target intensity pattern so that its periphery is artificially brightened—are often applied.

Beyond GS, alternative algorithms have been developed. NOVO-CGH (Zhang et al. 2017) directly optimizes a two-dimensional phase mask from a desired three-dimensional intensity distribution. More recently, Deep-CGH (Hossein Eybposh et al. 2020) employs convolutional neural networks (CNNs) to directly predict SLM phase masks from target intensity patterns. Because Deep-CGH can generalize once trained on a sufficiently large dataset of input–output pairs, and because feedforward prediction is substantially faster than iterative GS or NOVO-CGH optimization, this approach has the potential to become the dominant strategy in future applications.

To briefly summarize the opsins commonly used for two-photon optogenetic stimulation: among excitatory variants, *ChRmine* (Marshel et al. 2019) exhibits the largest photocurrents and highest light sensitivity; its red-shifted derivative *rsChRmine* (Kishi et al. 2022) minimizes crosstalk with imaging lasers; *Chronos* (Ronzitti et al. 2017) produces similarly large currents but with faster kinetics than ChRmine; *ChroME* (Klapoetke et al. 2014) is a modified form of Chronos; and *ChroME2s* (Sridharan et al. 2022) is a further engineered variant with enhanced performance. For inhibitory control, *GtACR1* (Govorunova et al. 2015) and its modified form *eGtACR1* (Mardinly et al. 2018) are frequently employed. More recently, potassium-selective opsins that are excitable by two-photon illumination (Govorunova et al. 2022; Vierock et al. 2022) have been developed, representing a promising direction for future research. In most cases, these opsins are targeted to the soma by inclusion of the Kv2.1 trafficking sequence.

It is well recognized that achieving balanced co-expression of opsins and calcium indicators in the same neurons via viral vectors is technically challenging. To overcome this, genetically engineered mouse lines such as Ai203 (TITL-st-ChroME-GCaMP7s-ICL-nls-mRuby3-IRES2-tTA2) (Bounds et al. 2023), which enable stable co-expression of opsins and GCaMP, have been developed. Similar lines may become the new standard for two-photon all-optical physiology in the near future.

Several recent studies have demonstrated the feasibility of two-photon optogenetic stimulation in deep brain structures. Vinograd et al. (Vinograd et al. 2024) perturbed attractor dynamics in the hypothalamus that encode emotional states and examined their stability. They used *Kv2.1-ChRmine* as the opsin and the NeuraLight 3D system (Bruker) for two-photon stimulation. The stimulation laser was a Monaco-1035-40-40 (1 MHz, 40 W, 300 fs), and the SLM was a 512 × 512 device, which is likely the standard configuration of the NeuraLight 3D.

Piantadosi et al. (Piantadosi et al. 2024) investigated the basolateral amygdala, identifying two neuronal populations associated with appetitive and aversive emotional behaviors. Selective stimulation of one ensemble suppressed behavior associated with the other. They also employed *Kv2.1-ChRmine* with a Bruker two-photon system, most likely the NeuraLight 3D given the reported 512 × 512 SLM. The stimulation laser was a Spirit One (1040 nm, ~8 W, 1 MHz, ~500 fs).

To this point, all implementations have been restricted to head-fixed animals. However, head fixation is a significant limitation for studying naturalistic behavior. A system termed 2P-FENDO (Accanto et al. 2023) enables the combined use of two-photon imaging and two-photon optogenetic stimulation in freely moving mice. Opsin expression relied on *Kv2.1-ChRmine*. The stimulation source was a Goji laser (Amplitude; 10 MHz, 8 W, 150 fs). The SLM was a Hamamatsu X10468-07 (792 × 600 pixels, 20 μm pitch, 15.8 × 12 mm, rise time 10 ms, fall time 80 ms), positioned at a Fourier plane. The distal end of a fiber bundle and the focal planes of GRIN lenses on both sides were conjugated to the same focal plane. Each fiber in the bundle (FIGH-15-600N, Fujikura) differs in optical path length by ~0.1 mm per meter, introducing relative delays on the order of 1 picosecond between fibers. Consequently, although a 100 femto second pulse is preserved within a single fiber, output from different fibers does not recombine temporally. Instead, the SLM generates a two-dimensional hologram corresponding to the physical arrangement of fibers, each mapped to the spatial position of target neurons. Because there is no crosstalk between fibers, this 2D holographic pattern is faithfully transferred to the GRIN lens focal plane. For example, if a single neuron corresponds to ~3 fibers, each laser pulse delivers three temporally separated stimuli to the same cell (**Fig. 2E**). It is likely that the fiber dimensions were deliberately matched to the size of target neurons to optimize spatial resolution.

## 5. Applying Multiphoton Techniques to BMI

Finally, we highlight brain–machine interfaces (BMIs) as an application that lies along the trajectory of two-photon optogenetics. BMIs map neural activity to the control of external devices (e.g., robotic arms or computer cursors), directly translating a subject's intentions into actions; they thus constitute a clinically important avenue with potential to improve the quality of life of patients with ALS or severe paralysis. Traditionally, neural activity has been recorded electrophysiologically, but single-cell recordings require invasive electrode implantation, and electrical stimulation co-activates passing axons, limiting spatial specificity (Histed, Bonin, and Reid 2009). Optical approaches offer a promising alternative (Ersaro, Yalcin, and Muller 2023; Hira 2024; Hira et al. 2014). Two-photon calcium imaging far exceeds electrophysiology in the number of simultaneously recorded neurons, and two-photon optogenetic stimulation remains the only method for precisely manipulating targeted ensembles. Here, we briefly outline examples of BMIs implemented with two-photon microscopy and sketch future directions informed by two-photon optogenetics. While some issues remain at the level of thought experiments, these technical challenges represent reasonable near-term goals for optics, electronics, and molecular biology; for a comprehensive discussion, see our previous review article (Hira 2024).

### 5-1. Two-Photon BMIs: Examples

The basic principle of BMI is to map neural signals from awake, behaving brains onto external actions; when the resulting outcomes are beneficial, subjects can adaptively modulate their brain activity. The central technical challenge is real-time readout. Under two-photon imaging, this requires integrating acquisition with online analysis and feedback. We built such a BMI in mouse primary/secondary motor cortex: using two-photon calcium imaging, we analyzed the activity of a single neuron in real time and coupled it to reward. Mice significantly increased the activity of the target cell within 15 minutes (Hira et al. 2014), indicating that reinforcement or attenuation of cortical activity synchronized to reward timing can be reproduced even under artificial manipulation. Furthermore, Mitani et al. (Mitani, Dong, and Komiyama 2018) developed a two-photon BMI targeting inhibitory cell types (PV, SOM, VIP) and demonstrated cell-type-dependent differences in controllability. In nonhuman primates, Trautmann et al. (Trautmann et al. 2021) monitored dendritic activity in macaque cortex with two-photon calcium imaging and successfully decoded movement intention in real time—an important step toward human application.

Taken together, BMIs based on two-photon microscopy already have a history spanning more than a decade, and, technically, extension to humans is increasingly within reach.

### 5-2. Future BMI based on two-photon optogenetics

Below, we consider a scenario in which two-photon microscopy propels BMI technology beyond electrophysiological measurement-and-control frameworks. To date, two-photon BMIs—including our own—have operated external devices with the activity of at most a few neurons. This scale can readily be expanded to hundreds, and, with large FOV two-photon microscopy, tens of thousands of neurons can be recorded over months (Hira, Townsend, et al. 2025). Thus, an increase in throughput by ~$10^4\times$ is already within reach. Learning has been shown to accelerate when two-photon BMI is paired with simultaneous one-photon photostimulation feedback (Prsa, Galiñanes, and Huber 2017); replacing this with two-photon optogenetic stimulation of targeted ensembles should further improve efficiency. Closed-loop experiments in which two-photon optogenetic stimulation is adjusted in real time based on two-photon calcium imaging have also been reported (Zhang et al. 2018). To stimulate novel ensembles in real time and in parallel, SLM phase masks must be generated on the fly; current SLMs can update within ≤2 ms, and DeepCGH can compute phase patterns on comparable timescales (Hossein Eybposh et al. 2020). As reviewed here, the scale of simultaneous two-photon stimulation looks promising with 3D-SHOT (Mardinly et al. 2018; Pégard et al. 2017) and its adaptation to large FOV systems (Fişek et al. 2023). Real-time image processing and high-speed analysis of large imaging streams during BMI operation are technically tractable using FPGAs, GPUs, or both (Shang et al. 2024). For use away from the bedside, miniaturization of two-photon microscopes will be necessary; apart from the laser, significant progress has already been made (Accanto et al. 2023).

In addition, optical methods now allow accurate in vivo measurement of neuromodulators including dopamine, acetylcholine, noradrenaline, serotonin, histamine, and opioids and, in some cases, optical control of their concentrations (Muir, Anguiano, and Kim 2024). Human translation of optogenetics will require safer viral-vector strategies, but progress is evident in studies of the retina and diseased human brain tissue (Kleinlogel et al. 2020; Ma et al. 2023; Sahel et al. 2021), as well as in viruses capable of crossing the blood–brain barrier (Challis et al. 2022; Chan et al. 2017). Given the maturity of electrophysiology, transparent electrodes that permit coexistence of optical and electrical interfaces will be valuable (Ramezani et al. 2024). Because interventions will be highly invasive, the ethical dimensions require careful consideration (Cabrera and Weber 2023).

Taken together, leveraging the fundamental advantages of optical measurement and control and coupling multimodal readouts with single-cell–resolution control in real-time closed loop should enable flexible, multifunctional, and task-general optical BMIs. Since the brain's three putative learning rules operate on distinct timescales (~10 ms, ~100 ms, and ~1 s;(Hira 2024)), we anticipate that BMI behavior will likewise exhibit qualitative transitions across approximately these three temporal regimes.

## 6. Conclusion

Focusing on two-photon optogenetic stimulation, we have surveyed a set of optical techniques that continue to evolve. We identified two major directions for activating opsins via two-photon absorption: (i) shaping the excitation pattern to match cellular morphology and (ii) employing low-repetition-rate femtosecond pulsed lasers. The first direction has been realized through spiral scanning, temporal focusing, and computer-generated holography (CGH). At present, a widely adopted strategy is to generate, at the sample plane under the objective, a hologram smaller than a neuron and rotate it in a spiral with galvanometric mirrors. We also note that 3D-SHOT—combining holography with temporal focusing—currently holds the record for the largest number of cells activated simultaneously. These approaches are being applied with optical fibers and GRIN lenses, as well as in large two-photon microscopes such as 2P-RAM and DIesel2p, and can be integrated into closed-loop experimental systems to advance toward brain–machine interfaces (BMIs). Our team is pursuing this trajectory, aiming to expand the bidirectional communication bandwidth between the brain and AI systems. Although the timing of clinical translation remains uncertain, we have outlined specific technical challenges that must be addressed. Taken together, advances in these component technologies and their combinations point toward a future that, while once reminiscent of science fiction, is becoming technically attainable. In parallel with progress in AI, the continued development of these methods will have significance that extends beyond basic neuroscience.